\documentclass[a4paper]{article}

\usepackage{INTERSPEECH2022}
\usepackage{url}
\usepackage{cite}
\usepackage{color, colortbl}
\definecolor{LightCyan}{rgb}{0.88,1,1}
\usepackage{array,multirow,graphicx}
\usepackage{xcolor}
\usepackage{array}

\title{Analysis of Voice Conversion and Code-Switching Synthesis Using VQ-VAE}
\name{Shuvayanti Das, Jennifer Williams, Catherine Lai}
\address{
  Centre for Speech Technology Research, The University of Edinburgh, UK \\
  Dept. of Electronics and Computer Science, University of Southampton, UK}
  
\email{shuvayanti@gmail.com, j.williams@soton.ac.uk, c.lai@ed.ac.uk}

\begin{document}

\maketitle
%

\begin{abstract}
This paper presents an analysis of speech synthesis quality achieved by simultaneously performing voice conversion and language code-switching using multilingual VQ-VAE speech synthesis in German, French, English and Italian. In this paper, we utilize VQ code indices representing phone information from VQ-VAE to perform code-switching and a VQ speaker code to perform voice conversion in a single system with a neural vocoder. Our analysis examines several aspects of code-switching including the number of language switches and the number of words involved in each switch. We found that speech synthesis quality degrades after increasing the number of language switches within an utterance and decreasing the number of words. We also found some evidence of \textit{accent transfer} when performing voice conversion across languages as observed when a speaker's original language differs from the language of a synthetic target utterance. We present results from our listening tests and discuss the inherent difficulties of assessing accent transfer in speech synthesis. Our work highlights some of the limitations and strengths of using a semi-supervised end-to-end system like VQ-VAE for handling multilingual synthesis. Our work provides insight into why multilingual speech synthesis is challenging and we suggest some directions for expanding work in this area. 
\end{abstract}

\section{Introduction}
Multilingual speech processing, such as code-switching in text-to-speech (TTS) \cite{zhang2019learning} and multilingual automatic speech recognition (ASR) \cite{toshniwal2018multilingual}, have gained a lot of momentum in recent years due to advances in machine learning. In particular, neural vocoders such as WaveNet \cite{oord2016wavenet} and WaveRNN \cite{kalchbrenner2018efficient} have made multilingual synthesis more feasible. In this paper we provide a detailed analysis of code-switching capabilities by introducing a methodology that is capable of synthesizing multilingual speech combined simultaneously with speaker voice conversion. Our technique introduces a new way to create speech that has observable non-native accents. 


Numerous speech processing applications stand to benefit from advances in high-quality multilingual speech synthesis, especially when techniques allow for controllability. 
One major application of speech synthesis is screen readers for the visually impaired population. Screen readers generate speech out loud from the text found on webpages, allowing visually impaired users to use computers effectively. Most screen readers are limited to single languages and this creates a problem as users browse the multilingual internet \cite{rodriguez2015exploring}.

Even still, it is not enough to create multiple TTS systems for every language as this is not an efficient way to address gaps in code-switching synthesis capabilities. Our work in this paper addresses the gap in multilingual synthesis by providing a technique for fine-grained code-switching combined with multi-speaker voice conversion. The samples that we analyze in this paper do not require specialized multilingual corpora: all of the data from multiple languages is combined together during the model training. 

We are the first to explore code-switching and voice conversion in terms of the number of language switches and granularity of linguistic units (i.e., number of words) in each switch. We also introduce a new aspect of synthesis called \textit{accent transfer} where we generate speech that sounds like a non-native speaker. To the best of our knowledge, we are the first to conduct this analysis using discrete phone and speaker representations from representations learned with VQ-VAE~\cite{oord2017neural,williams21_ssw} and synthesized with the WaveRNN~\cite{kalchbrenner2018efficient} neural vocoder. We are also the first to propose using a multilingual VQ-VAE system for the purpose of evaluating accent transfer in cross-language voice conversion combined with code-switching. Our contributions are:
\begin{enumerate}
    \item Evaluate code-switching synthesis quality in four languages based on number of language switches
    \item Evaluate code-switching synthesis quality as above, combined with differing granularity of the phrases during code-switching (number of words per switch) 
    \item New technique to synthesize and evaluate non-native accented speech
\end{enumerate}

\section{Related Work}


Most existing work has treated multilingual speech synthesis independently of voice conversion. However, very recent work from \cite{zhang2019learning} describes how to use Tacotron2 with speaker embeddings to create fluent-sounding speech in one language, using the voice of a different speaker from another language. For example, the voice from a monolingual English speaker can `speak' fluently in Spanish. Their trained model had capabilities in English, Spanish and Chinese. Their technique of combining data from all languages together for model training is similar to the model that we use in our paper. Importantly, the work that we present in our paper involves multilingual \textit{code-switching} which allows for languages to change within a single utterance. 

The work of \cite{oord2017neural} was the original VQ-VAE that defined a fully self-supervised autoencoder system to learn intermediate representations of audio. The VQ-VAE system has three main components: encoder, VQ clustering, and decoder. Embeddings are learned during the VQ clustering process and they represent the content of speech, similar to sub-phone units because they behave like phones but there are more types than a traditional phoneset. The original VQ-VAE was re-formulated as a dual-encoder system first by \cite{zhao2020improved} (phones and F0) and then by \cite{williams2021learning} (phones and speaker). Our work is based on the dual-encoder system from \cite{williams2021learning} which models the content (as a sequence of VQ phone codes) as well as speaker (as a single VQ speaker code). Finally \cite{williams21_ssw} further expanded the dual-encoder model from \cite{williams2021learning} by showing it is possible to convert a monolingual multi-speaker model trained on VCTK data \cite{yamagishi2019cstr} to a multilingual multi-speaker model trained on four languages combined together from the SIWIS dataset \cite{267a8b2712774f4aa634e45af118c711}. 

Although VQ-VAE in our experiments does not deal directly with text for TTS synthesis, the work of \cite{yasuda2021end} showed that VQ-VAE is well-suited for TTS. Their technique learned a mapping of graphemes into discrete VQ phone code indices, and then they provided the VQ phone codes to a trained neural vocoder for speech synthesis. It was also shown by \cite{fong21b_ssw} that sequences of VQ phone codes can be taken out of context and re-assembled through concatenation, similar to unit-selection synthesis. 
They determined that speaking rate was a factor in speech intelligibility. In this paper, we also experiment with speaking rate and use an approach similar to `unit synthesis' to simulate code-switching with VQ phone codes. Ours is the first study to examine this capability using VQ-VAE from multi-speaker and multi-language data. 

Accent is often something that researchers seek to control or model for the purpose of removing non-native accents from speech or measuring it for pronunciation scoring~\cite{minh2021system, fu2022improving, lin21j_interspeech}. Recent work in \cite{zhao2019foreign} presents a method to effectively remove a non-native accent in order to make a speaker sound more fluent in the target language. Recent work in \cite{wang2021accent} uses speech disentanglement to change dialectal accent within a single language: Chinese. They used an encoder-decoder model with adversarial training and data from two distinct Chinese dialects. They evaluated accentedness of synthetic speech using an A/B test, measuring listener judgements for the similarity of each sample to a reference. In our work in this paper, we explore accent transfer across languages, for example, making an English utterance sound as if it was spoken by a French speaker. We also introduce a new technique for accent evaluation based on perceived proficiency levels.

\section{Methodology}
We examine language mixing in terms of the number of times the languages change within an utterance, the number of words involved in each switch, and controlling the speaker's voice (the multi-language data that we use for code-switching is also multi-speaker). Our experiments are based on speech that we generated using a freely available pre-trained VQ-VAE + WaveRNN model and architecture from \cite{williams21_ssw}\footnote{\url{https://github.com/rhoposit/multilingual_VQVAE}}. The VQ-VAE phone encoder encodes raw audio into a sequence of embeddings represented by a sequence of discrete indices (\textit{phone codes}) in a look-up table whereas speakers are encoded as a single embedding and discrete index (\textit{speaker code}) from a separate table. 

\begin{figure}[ht]
  \centering
  \includegraphics[width=0.5\textwidth]{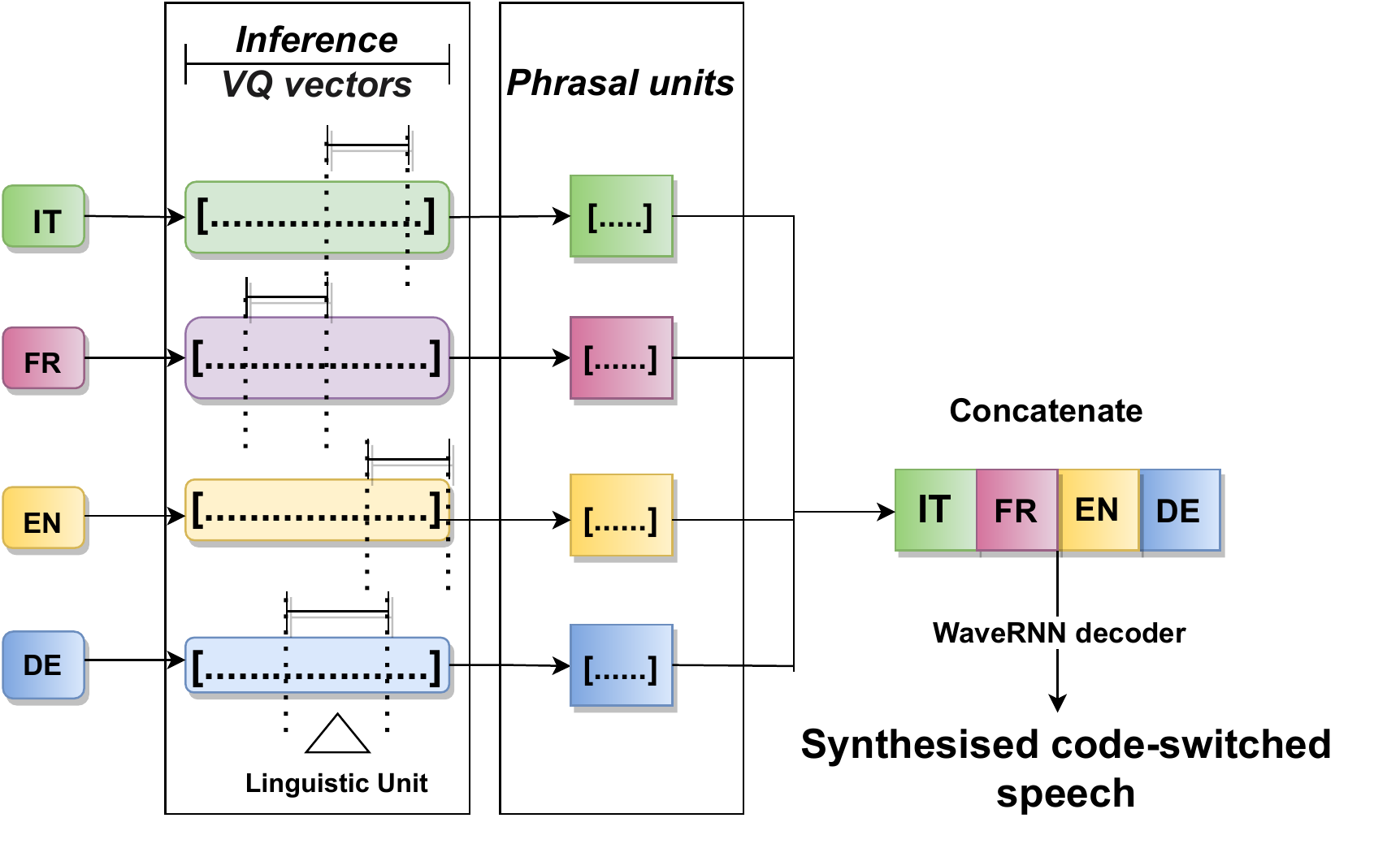}
  \caption{Example of a code-switched utterance using phone codes from VQ-VAE for multiple languages.}
  \label{units-example}
  \vspace{-4mm}
  \end{figure}
  
One of the benefits of using the multilingual dual-encoder VQ-VAE architecture is that it allows us to represent and control content information and speaker information separately but also simultaneously within the same system with one neural vocoder. The pre-trained model that we used in this paper included a one-hot global language ID code which was used for the WaveRNN decoder which \cite{williams21_ssw} used to stabilize the model training. In our experiments, we set always set the global language code to be the first language in a code-switched utterance. 

As the pre-trained model that we use in this work was trained on SIWIS~\cite{267a8b2712774f4aa634e45af118c711} data in English, French, German and Italian, we also used this dataset in our code-switching experiments. It contains 36 speakers who are at least bilingual, though some speak more languages. There is a balance across genders, though the languages are skewed slightly toward French being most frequent followed by English. We utilized samples from the dataset corresponding with the train split from \cite{williams21_ssw} in order to focus our analysis on code-switching granularity rather than model generalization. 

\subsection{Controlling Language Mixtures}
We used two different methods to control the language mixtures: 1) concatenating sequences of VQ phone codes from whole utterances in different languages, and 2) concatenating VQ phone codes from randomly selected words in different languages. The VQ phone codes were obtained by running an inference step using the pre-trained VQ-VAE model. Audio from wav files is provided as input to the VQ-VAE system and the resulting VQ embeddings are saved from each lookup table. For the VQ phone codes, this is a sequence of embeddings and discrete indices per-utterance, and for the VQ speaker codes, this is a single embedding and discrete index per-utterance.  

For the first set of code-switching experiments, we concatenated VQ phone codes from entire files in different languages, to form very long utterances. We used these to evaluate speech synthesis quality from only the number of language switches. Our second set, we simulated a more realistic code-switching scenario where consecutive words were randomly selected from utterances and the corresponding VQ phone codes were concatenated together. To determine which phone codes corresponded to words, we first obtained forced alignments using multilingual CLARIN BAS~\cite{kisler2017multilingual} and calculated the alignment between VQ phone codes and word-level timestamps based on the VQ-VAE downsampling factor where: 1000 ms of speech equals 427 VQ codes. The process of concatenating VQ phone codes from multiple languages is shown in Figure~\ref{units-example}.

\subsection{Controlling Speaker Voice}
VQ speaker codes for single-speaker audio files are obtained during inference at the same time as VQ phone codes. Speaker VQ codes are treated as global conditions when passed to the WaveRNN decoder. Therefore the speaker can be controlled by changing the code on a per-utterance basis. Our experiments use a single speaker code even when synthesizing a mixed-language utterance with data concatenated from multiple different speakers from different languages.

\section{Experiments and Results}
In this section, we describe the outcomes for three experiments: language code-switching with full utterances, language code-switching with word units, and accent transfer. All of our code-switched and voice-conversion speech samples were evaluated by human listeners. Participants in our listening study were recruited from Prolific and paid the equivalent of \pounds 7.50 per hour. Our survey was administered via Qualtrics. Each experiment was evaluated slightly differently, described in more details below. We encourage the reader to listen to our audio samples\footnote{\url{https://rhoposit.github.io/interspeech2022/}}.

  \vspace{-2mm}
\subsection{Language Code-Switching}
  \vspace{-3mm}
We examined the quality of code-switched speech by varying the number of \textit{language switches} as well as the \textit{size of linguistic unit} within an single utterance. We combined VQ phone codes from speech in the following languages to simulate code-switching: \textbf{English}, \textbf{French}, \textbf{German}, and \textbf{Italian}. We varied the number of language switches within an utterance in the range of $[4,8,12]$ in any order and the languages were chosen at random (the same language did not occur consecutively). For experiments using full utterances, we randomly selected utterances from the target languages, obtained the VQ phone codes, and concatenated the VQ phone codes together before the final re-synthesis step. For experiments involving linguistic units, we randomly selected words based on forced alignment boundaries, obtained the VQ phone codes, and concatenated the VQ phone codes together before the final re-synthesis step. For the unit-based experiments, we varied the size of linguistic unit as the number of consecutive words using values of $[4,8,12]$. The number of words per linguistic unit is intended to mimic short versus long phrases. 
  \vspace{-2mm}
\subsection{Results for Language Code-Switching}
  \vspace{-3mm}
All of the samples were evaluated with a listening test where participants were asked to listen to 18 utterances (for the full utterance concatenation) and 54 utterances (for the word-based concatenation). 20 participants were recruited who were at least trilingual, as determined with a Prolific pre-screening survey. Listeners were asked to rate naturalness and intelligibility on a Likert scale of 1-5 (where 5 is most natural/intelligible). They were also asked to determine the number of language switches using a categorical choice (\textit{4}, \textit{8}, or \textit{12}) and likewise to determine the length of word-units using a categorical choice (\textit{4}, \textit{8}, or \textit{12}). Participants were provided with samples in a randomized order.

We calculated a mean-opinion scores (MOS) for naturalness and intelligibility ratings. We grouped results based on the speaker speed (calculated as avg. words per second for all utterances from a given speaker). Slow speakers were identified as speaking fewer than $3.5$ words per second on average. The group of slow speakers had an average speaking rate of $2.9$ words/sec and were the slowest speakers in the dataset, IDs: 19(M), 24(F), 26(M), 29(F), and 33(M). The group of fast speakers had an average speaking rate of $3.6$ words/sec, IDs: 18(F), 25(F), and 35(M).


\begin{table}[ht!] 
\centering
\caption{EXP \#1: Naturalness and intelligibility MOS scores for slow and fast speakers, based on number of language switches within an utterance.}
    \resizebox{0.42\textwidth}{!}{\begin{tabular}{|c|c|c|c|}
    \hline
Speaker & Language & \multicolumn{2}{c|}{MOS} \\
      Speed & Switches & Naturalness & Intelligibility \\
      \hline
      & 4 & 3.2 & 4.2 \\
       Slow & 8 & 3.3 & 3.4 \\
       & 12 & 3.5 & 4.2 \\ \hline
       \multicolumn{2}{|r|}{Avg. Slow} & \textbf{3.3} & \textbf{3.9} \\ \hline
       & 4 & 2.8 & 3.4 \\
        Fast & 8 & 2.7 & 3.0 \\
       & 12 & 3.1 & 3.2 \\ \hline
       \multicolumn{2}{|r|}{Avg. Fast} & \textbf{2.8} & \textbf{3.2} \\ \hline
    \end{tabular}}
\label{mos_int_switches}
\end{table}

\begin{table}[h!] 
    \centering
\caption{EXP \#2: Naturalness and intelligibility MOS for slow and fast speakers, showing number of language switches and number of words per unit (within each language segment). Results for highest intelligibility are highlighted per category. }    
    \resizebox{0.46\textwidth}{!}{\begin{tabular}{|c||c|c|c|c|}
      \hline
      &  Number & Language & \multicolumn{2}{c|}{MOS}\\
      &  Words & Switches & Naturalness & Intelligibility \\
      \hline
      \parbox[t]{2mm}{\multirow{9}{*}{\rotatebox[origin=c]{90}{Slow}}} & \multirow{3}{*}{4} & 4 & 3.7 & 3.1  \\
      & & 8 & 3.5	& 3.3\\
      & &  \cellcolor{LightCyan}12 & \cellcolor{LightCyan}3.6	& \cellcolor{LightCyan}3.8 \\
      \cline{2-5}
      & \multirow{3}{*}{8} & 4 & 3.9 & 3.6  \\
      & & \cellcolor{LightCyan}8 & \cellcolor{LightCyan}3.6 & \cellcolor{LightCyan}4.1\\
      & & 12 & 3.1	& 3.5 \\
      \cline{2-5}
      & \multirow{3}{*}{12} & 4 & 3.6 & 3.2\\
      & & 8 & 3.4 & 2.4\\
      & & \cellcolor{LightCyan}12 & \cellcolor{LightCyan}2.1	& \cellcolor{LightCyan}3.6 \\
      \hline\hline
      \parbox[t]{2mm}{\multirow{9}{*}{\rotatebox[origin=c]{90}{Fast}}} & \multirow{3}{*}{4}& \cellcolor{LightCyan}4 & \cellcolor{LightCyan}3.1 & \cellcolor{LightCyan}3.1\\
      & & 8 & 1.9 & 3.1\\
      & & 12 & 2.3& 2.9\\
      \cline{2-5}
      & \multirow{3}{*}{8}& 4 & 3.1 & 2.8\\
      & & 8 & 3.3 & 1.7\\
      & & \cellcolor{LightCyan}12 & \cellcolor{LightCyan}3.5 & \cellcolor{LightCyan}3.5\\
      \cline{2-5}
      & \multirow{3}{*}{12} & \cellcolor{LightCyan}4 & \cellcolor{LightCyan}3.6 & \cellcolor{LightCyan}3.7\\
      & & 8 & 3.8 & 3.9\\
      & & 12 & 3.3 & 1.9\\
      \hline
    \end{tabular}}
    \label{units-naturalness}
  \vspace{-4mm}
  \end{table}


\break
\noindent\textbf{EXP \#1: Full utterances, mixed languages.} Table~\ref{mos_int_switches} shows the MOS ratings for the number of language switches and distinguishes between slow and fast speakers. This table shows that on average the slow speakers have better naturalness and intelligibility compared to fast speakers. The speaker speed does not appear to impact intelligibility as much as naturalness, though the two measures are interrelated. The number of language switches within an utterance does not seem to have a negative impact on perceived naturalness or intelligibility, which is counter-intuitive but also a very promising finding for multilingual speech synthesis. 


\noindent\textbf{EXP \#2: Word-based units, mixed languages.}
Table~\ref{units-naturalness} shows the mean-opinion scores (MOS) for naturalness and intelligibility when considering both the number of words per language segment and the number of language switches. As before, synthesized speech from fast versus slow speakers are separated for comparison purposes. Among slow speakers, the naturalness decreases when the number of words and language switches is very high, while the intelligibility is generally preserved. On the other hand, among fast speakers the naturalness is generally preserved for high values of language switches and number of words, but the intelligibility suffers. 
When language segments are longer (such as 12) it is reasonable that this improves intelligibility because the utterances are more cohesive and meaningful for the listener to perceive. While a high amount of language switches (such as 12) can result in increased naturalness or intelligibility, very long audio is much more challenging to synthesize with the WaveRNN vocoder and will eventually test the limits of WaveRNN capabilities. 

\begin{figure}[t!]
  \centering
  \includegraphics[width=0.232\textwidth]{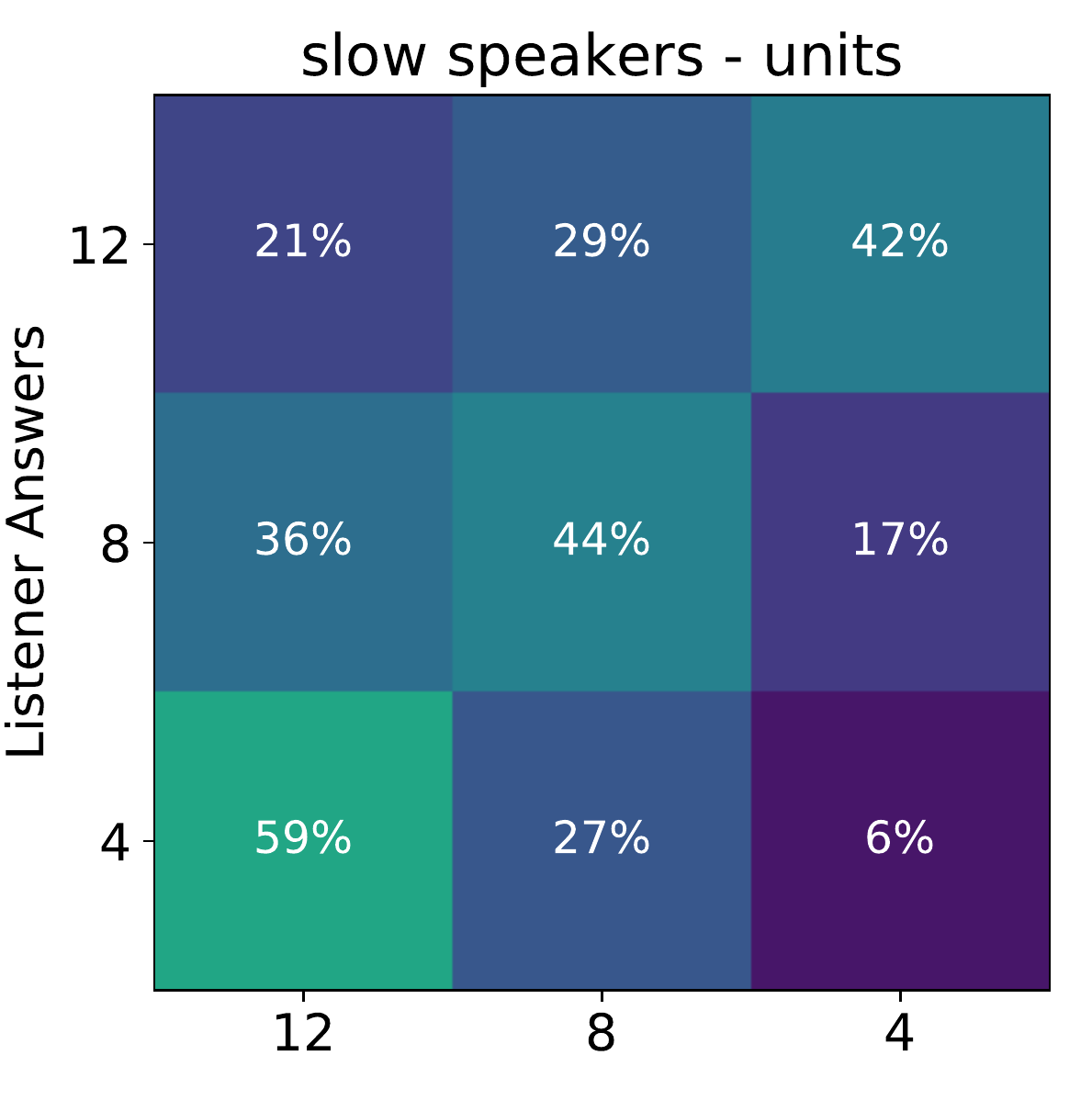}\includegraphics[width=0.25\textwidth]{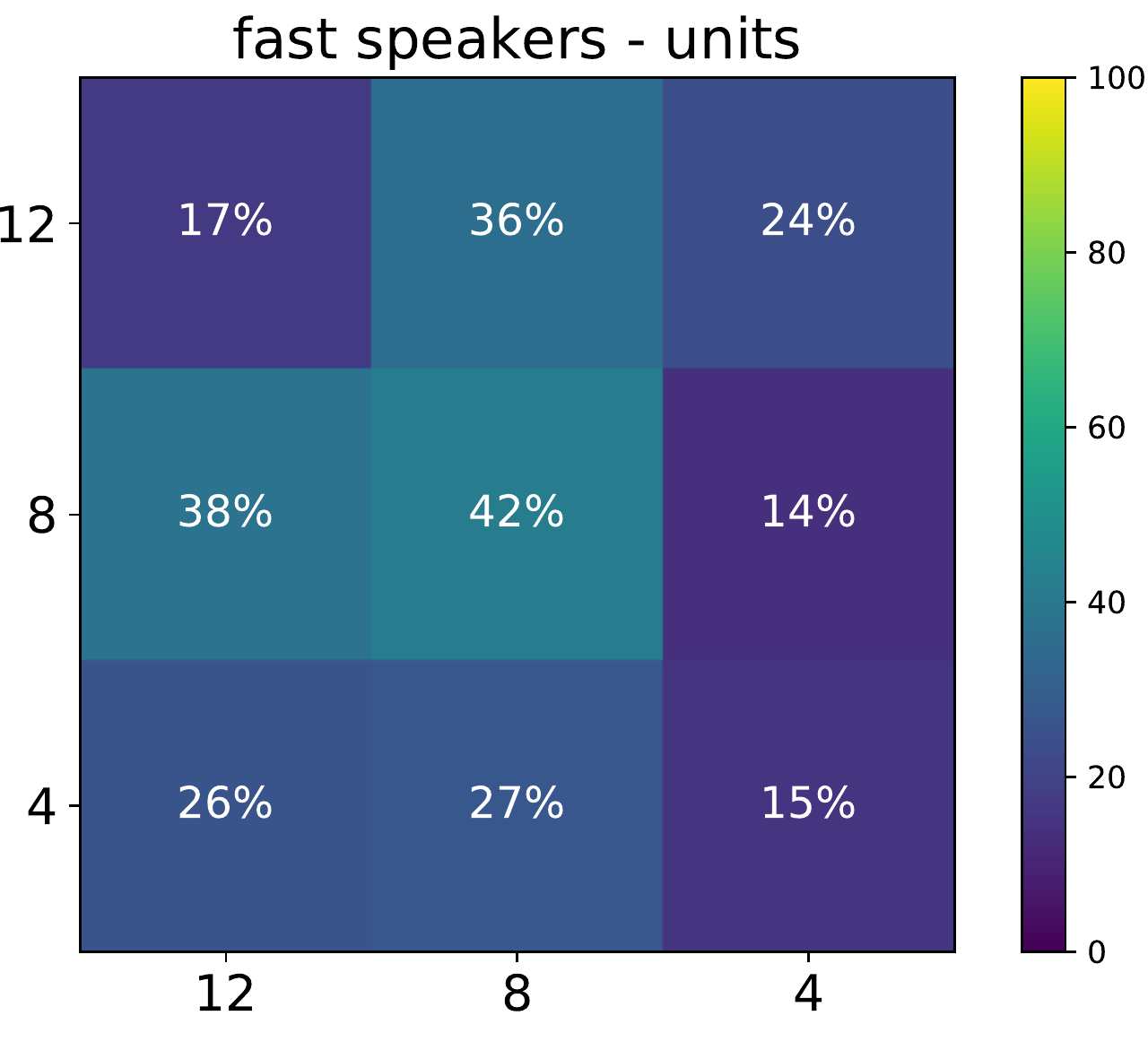}
  \includegraphics[width=0.23\textwidth]{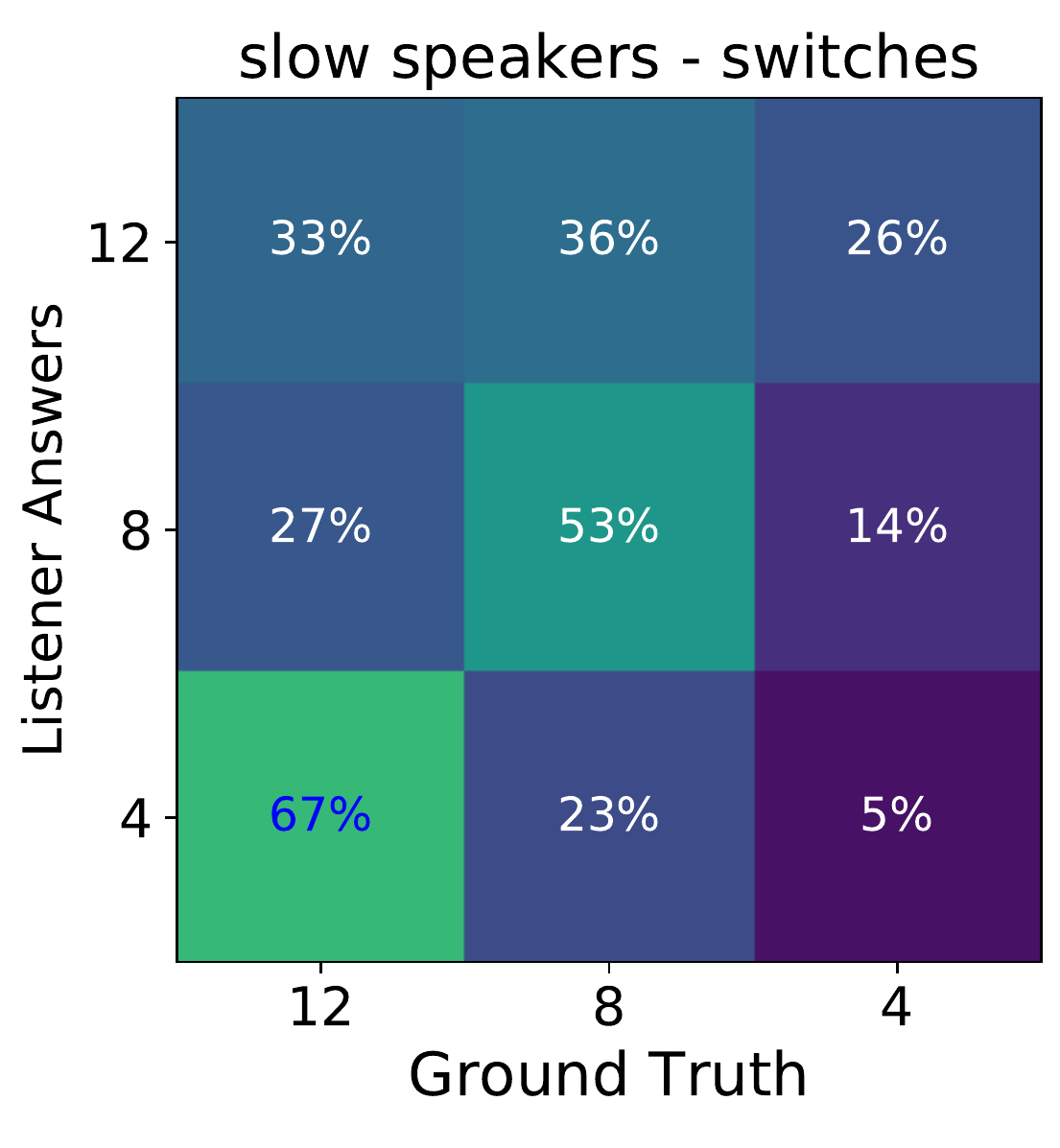}\includegraphics[width=0.252\textwidth]{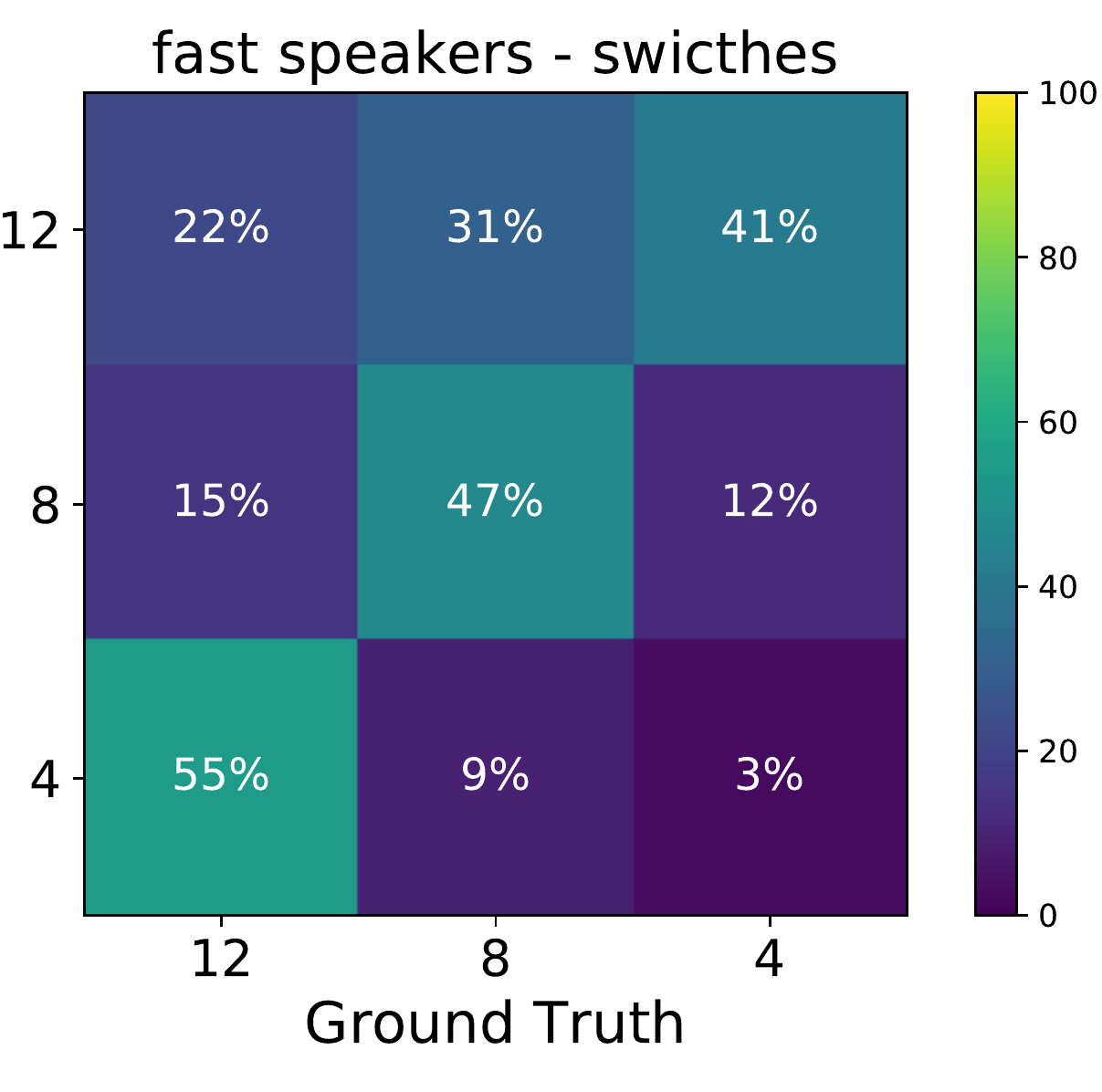}
  \caption{EXP \#2: Percentage of listeners who correctly guessed the number of language switches and unit sizes. }
  \label{confusion_matrix_2}
  \vspace{-5mm}
\end{figure}

We asked listeners to judge the number of language switches and the number of words per language segment and report the percent of listeners who answered correctly in Figure~\ref{confusion_matrix_2}. 
Overall, performance is better for slow speakers and judging the number of language switches was easier than the number of words per segment. To further characterise listener perceptions we calculated Pearson's correlation coefficient $\rho$ using the listener categorical judgements. We found $\rho = 0.78$ (p-value $<$ 0.05) demonstrating that listener performance is influenced more by the language switches than the length of word units.

\begin{table}[ht!] 
    \centering
    \caption{Summary of perceived English and French proficiency trends for synthesized code-switched utterances from each speaker. The average represents the average speaker proficiency level as judged across all listeners.}
    \resizebox{0.48\textwidth}{!}{
    \begin{tabular}{||c|c|c|c|c|c|c|c|c||}
      \hline
      Speaker & Perceived & \multicolumn{6}{|c|}{Rating Frequency (\%)} & Avg. \\
      ID & Proficiency & \multicolumn{6}{|c|}{} &\\
      \hline
        & & 0 & 1 & 2 & 3 & 4 & 5 & \\
      \hline
      \hline
      \multirow{2}{*}{18 } & English & 19 &  35 & 25 & 13 & 8 & 0 & 1.6\\       
      & French & 2 & 0 & 6 & 27 & 46 &  19 & 3.7 \\
       \hline
       \multirow{2}{*}{19 } & English & 0 &  6 & 9 & 23 & 31 & 31 & 3.7\\
       & French & 6 & 15 & 21 & 19 & 25  & 14 & 2.9 \\
       \hline
        \multirow{2}{*}{35 } & English & 35 & 38 & 10 & 2 & 11 & 4 & 1.3\\
       & French & 0 & 6 & 6 & 21 & 46 & 21 & 3.7 \\
       \hline
        \multirow{2}{*}{25} & English & 0 & 10 & 15 & 40 & 27 & 8 & 3.1\\
        & French & 0 & 6 & 19 & 42 & 15 & 18 & 3.2 \\
       \hline
        \multirow{2}{*}{33} & English & 0 & 0 & 2 & 19 & 52 & 27 & 4.0\\
       & French & 4 & 19 & 19 & 25 & 25 & 8 & 2.0 \\
       \hline
       \multirow{2}{*}{29} & English & 11 & 24 & 22 & 23 & 16 & 4 & 2.2\\
       & French & 6 & 21 & 14 & 22 & 20 & 17 & 2.8 \\
       \hline
       \hline
    \end{tabular}}
    \label{MOS-trend-code-switched-utt}
    \vspace{-4mm}
\end{table}

\subsection{Accent Transfer from Cross-Lingual Voice Conversion}
We experimented with accent transfer while performing voice conversion across languages in code-switched utterances. We synthesized speech for a given speaker from the SIWIS dataset who did not have any speech samples from the target language. It is similar to the idea of making a particular speaker sound as if they are ``speaking'' in a foreign language with an accent. The languages used for these experiments were \textbf{English} and \textbf{French}. We tried different language orders because \cite{williams21_ssw} had reported that within-utterance speaker consistency varied slightly with language order. We carefully selected SIWIS speakers for accent transfer that were at least bilingual in any of the languages. We designed the synthesis samples such that the speaker would be artificially `speaking' English/French during English-French and French-English code-switching.

\subsection{Results from Accent Transfer}
\vspace{-2mm}

Listeners were native and bilingual English/French speakers. For example, in a code-switched sample of English-French, it was rated by a listener with an English mother-tongue and again by a listener with a French mother-tongue. The listeners were asked to rate the accent of the speakers in 14 samples to determine if they sound proficient using a Likert scale of 1-5 (where 3+ is fluent and 5 is native/bilingual). Table \ref{MOS-trend-code-switched-utt} reports the percentage of listeners who rated each speaker and target language. Speaker 19 (a bilingual speaker of French-Italian) shows some degradation in perceived French proficiency, which is likely due to accent transfer from English. Speaker 18 (a trilingual English-French-Italian speaker) shows degradation in English proficiency. Similar accent transfer was also observed for Speaker 35 who has English as a bilingual language.  

We used the proficiency ratings to determine if changing the order of the languages had any effect on the listener perception speaking proficiency. For Speakers 19, 35 and 29, we found a Pearson correlation of $\rho = 0.44$, $\rho = 0.57$ and $\rho = 0.52$ respectively (p-value $<$ 0.05) indicating a slight correlation between language order and listener perception of accent. This shows that the position of the first language has a small effect on the speaker's perceived `foreign' accent. Overall, a Pearson correlation coefficient of $\rho = 0.73$ (p-value $<$ 0.05) indicates that the leading language influenced the perceived speaker language proficiency in both English and French.

\section{Discussion}
We have shown how code-switched speech is perceived by listeners when we vary the number of languages and length of segment during code-switching. Listeners were more influenced by frequent language switches. From our accent transfer experiments, slow speakers were more likely to be perceived as native/bilingual proficiency compared to fast speakers. We discovered that listener perception of speaking fluency was influenced by the leading language during code-switching which could be a consequence of the global language code in the WaveRNN vocoder. Future work should explore the reasoning behind why listeners perceive slower speakers to be native/bilingual and whether this changes based on different types of vocoders. The work we have presented in this paper contributes important insights for the continuing development of multilingual speech applications.


\bibliographystyle{IEEEtran}

\bibliography{mybib}

\begin{thebibliography}{10}
\providecommand{\url}[1]{#1}
\csname url@samestyle\endcsname
\providecommand{\newblock}{\relax}
\providecommand{\bibinfo}[2]{#2}
\providecommand{\BIBentrySTDinterwordspacing}{\spaceskip=0pt\relax}
\providecommand{\BIBentryALTinterwordstretchfactor}{4}
\providecommand{\BIBentryALTinterwordspacing}{\spaceskip=\fontdimen2\font plus
\BIBentryALTinterwordstretchfactor\fontdimen3\font minus
  \fontdimen4\font\relax}
\providecommand{\BIBforeignlanguage}[2]{{%
\expandafter\ifx\csname l@#1\endcsname\relax
\typeout{** WARNING: IEEEtran.bst: No hyphenation pattern has been}%
\typeout{** loaded for the language `#1'. Using the pattern for}%
\typeout{** the default language instead.}%
\else
\language=\csname l@#1\endcsname
\fi
#2}}
\providecommand{\BIBdecl}{\relax}
\BIBdecl

\bibitem{zhang2019learning}
Y.~Zhang, R.~J. Weiss, H.~Zen, Y.~Wu, Z.~Chen, R.~Skerry-Ryan, Y.~Jia,
  A.~Rosenberg, and B.~Ramabhadran, ``{Learning to Speak Fluently in a Foreign
  Language: Multilingual Speech Synthesis and Cross-Language Voice Cloning},''
  \emph{Proceedings of Interspeech 2019}, pp. 2080--2084, 2019.

\bibitem{toshniwal2018multilingual}
S.~Toshniwal, T.~N. Sainath, R.~J. Weiss, B.~Li, P.~Moreno, E.~Weinstein, and
  K.~Rao, ``{Multilingual Speech Recognition with a Single End-to-End Model},''
  in \emph{2018 IEEE international conference on acoustics, speech and signal
  processing (ICASSP)}, 2018, pp. 4904--4908.

\bibitem{oord2016wavenet}
A.~{van den Oord}, S.~Dieleman, H.~Zen, K.~Simonyan, O.~Vinyals, A.~Graves,
  N.~Kalchbrenner, A.~Senior, and K.~Kavukcuoglu, ``{WaveNet: A Generative
  Model for Raw Audio},'' in \emph{Proceedings of the 9th ISCA Speech Synthesis
  Workshop (SSW9), Sunnyvale, USA}, 2016, p. 125.

\bibitem{kalchbrenner2018efficient}
N.~Kalchbrenner, E.~Elsen, K.~Simonyan, S.~Noury, N.~Casagrande, E.~Lockhart,
  F.~Stimberg, A.~van~den Oord, S.~Dieleman, and K.~Kavukcuoglu, ``{Efficient
  Neural Audio Synthesis},'' in \emph{International Conference on Machine
  Learning, Stockholm, Sweden}, 2018, pp. 2410--2419.

\bibitem{rodriguez2015exploring}
S.~Rodr{\'\i}guez~V{\'a}zquez, ``{Exploring Current Accessibility Challenges in
  the Multilingual Web for Visually-Impaired Users},'' in \emph{Proceedings of
  the 24th International Conference on World Wide Web}, 2015, pp. 871--873.

\bibitem{oord2017neural}
A.~van~den Oord, O.~Vinyals, and K.~Kavukcuoglu, ``{Neural Discrete
  Representation Learning},'' in \emph{Proceedings of the 31st International
  Conference on Neural Information Processing Systems (NIPS)}, 2017, pp.
  6309--6318.

\bibitem{williams21_ssw}
J.~Williams, J.~Fong, E.~Cooper, and J.~Yamagishi, ``{Exploring Disentanglement
  with Multilingual and Monolingual VQ-VAE},'' in \emph{Proceedings of the 11th
  ISCA Speech Synthesis Workshop (SSW11), Budapest, Hungary}, 2021, pp.
  124--129.

\bibitem{zhao2020improved}
Y.~Zhao, H.~Li, C.-I. Lai, J.~Williams, E.~Cooper, and J.~Yamagishi,
  ``{Improved Prosody from Learned F0 Codebook Representations for VQ-VAE
  Speech Waveform Reconstruction},'' in \emph{Proceedings of Interspeech 2020,
  Shanghai, China}, 2020, pp. 4417--4421.

\bibitem{williams2021learning}
J.~Williams, Y.~Zhao, E.~Cooper, and J.~Yamagishi, ``{Learning Disentangled
  Phone and Speaker Representations in a Semi-Supervised VQ-VAE Paradigm},'' in
  \emph{Proceedings of 2021 IEEE International Conference on Acoustics, Speech
  and Signal Processing (ICASSP)}, 2021, pp. 7053--7057.

\bibitem{yamagishi2019cstr}
J.~Yamagishi, C.~Veaux, K.~MacDonald \emph{et~al.}, ``{CSTR} {VCTK} {C}orpus:
  {E}nglish {M}ulti-{S}peaker {C}orpus for {CSTR} {V}oice {C}loning {T}oolkit
  (version 0.92),'' \emph{Edinburgh DataShare}, 2019.

\bibitem{267a8b2712774f4aa634e45af118c711}
J.-P. Goldman, P.-E. Honnet, R.~Clark, P.~Garner, M.~Ivanova, A.~Lazaridis,
  H.~Liang, T.~Macedo, B.~Pfister, M.~Ribeiro, E.~Wehrli, and J.~Yamagishi,
  ``{The SIWIS Database: A Multilingual Speech Database with Acted Emphasis},''
  in \emph{Proceedings of Interspeech 2016}, 2016, pp. 1532--1535.

\bibitem{yasuda2021end}
Y.~Yasuda, X.~Wang, and J.~Yamagishi, ``{End-to-end Text-to-Speech Using Latent
  Duration Based on VQ-VAE},'' in \emph{Proceedings of 2021 IEEE International
  Conference on Acoustics, Speech and Signal Processing (ICASSP)}, 2021, pp.
  5694--5698.

\bibitem{fong21b_ssw}
J.~Fong, J.~Williams, and S.~King, ``{Analysing Temporal Sensitivity of VQ-VAE
  Sub-phone Codebooks},'' in \emph{Proceedings of the 11th ISCA Speech
  Synthesis Workshop (SSW11), Budapest, Hungary}, 2021, pp. 227--231.

\bibitem{minh2021system}
N.~Q. Minh and P.~D. Hung, ``{The System for Detecting Vietnamese
  Mispronunciation},'' in \emph{International Conference on Future Data and
  Security Engineering}.\hskip 1em plus 0.5em minus 0.4em\relax Springer, 2021,
  pp. 452--459.

\bibitem{fu2022improving}
K.~Fu, S.~Gao, K.~Wang, W.~Li, X.~Tian, and Z.~Ma, ``{Improving Non-native
  Word-level Pronunciation Scoring with Phone-level Mixup Data Augmentation and
  Multi-source Information},'' \emph{arXiv preprint arXiv:2203.01826}, 2022.

\bibitem{lin21j_interspeech}
B.~Lin and L.~Wang, ``{Deep Feature Transfer Learning for Automatic
  Pronunciation Assessment},'' in \emph{Proceedings of Interspeech 2021}, 2021,
  pp. 4438--4442.

\bibitem{zhao2019foreign}
G.~Zhao, S.~Ding, and R.~Gutierrez-Osuna, ``{Foreign Accent Conversion by
  Synthesizing Speech from Phonetic Posteriorgrams},'' in \emph{Proceedings of
  Interspeech}, 2019, pp. 2843--2847.

\bibitem{wang2021accent}
Z.~Wang, W.~Ge, X.~Wang, S.~Yang, W.~Gan, H.~Chen, H.~Li, L.~Xie, and X.~Li,
  ``{Accent and Speaker Disentanglement in Many-to-many Voice Conversion},'' in
  \emph{2021 12th International Symposium on Chinese Spoken Language Processing
  (ISCSLP)}, 2021, pp. 1--5.

\bibitem{kisler2017multilingual}
T.~Kisler, U.~Reichel, and F.~Schiel, ``{Multilingual Processing of Speech via
  Web Services},'' \emph{Computer Speech \& Language}, vol.~45, pp. 326 -- 347,
  2017.

\end{thebibliography}

\end{document}